\documentclass[aps,prb,preprint,floatfix]{revtex4-1}
\usepackage{amsmath,amssymb,epsfig}

\begin{document}

\title{Exact ferromagnetic ground state of pentagon chains}

\author{Miklos~Gul\'acsi, Gy\"orgy~Kov\'acs, 
and Zsolt~Gul\'acsi}
\address{
Department of Theoretical Physics, University of Debrecen, 
H-4010 Debrecen, Hungary}

\date{June 21, 2014}

%
%
%
%
%

\begin{abstract}
We model conducting pentagon chains with a multi orbital Hubbard
model and prove that well below half filling exact ferromagnetic
ground states appear. The rigorous method we use is based on the
transformation of original hamiltonian into positive semidefinite 
form. This technique is independent of the spatial 
dimesion and does not require integrability of the model.
The obtained ferromagnetism is connected to dispersionless bands 
but in a much broader sense than flat band ferromagnetism requires, 
where on every site a Hubbard term is present. 
In our case only a small percentage of,
even randomly distributed, sites are only interacting. 
\end{abstract}

\maketitle

Recent observation of ferromagnetism in polythiophene compounds \cite{one} 
has generated a widespan interest and a heightened research
effort to develop plastic ferromagnets and more generally to 
understand ferromagnetism in systems made entirely of nonmanetic
elements. As a result several theories have emerged to describe 
ferromagnetism in these systems \cite{ketto}, with particular focus
on ferromagnetism due to dispersionless bands \cite{harom,haroma} or periodic
Anderson model \cite{negy}. 

These theories however, are centered on two particle descriptions, 
on exact diagonalizations of small samples, or mean field 
type of approximations \cite{MF}, i.e., exploring 
weak interacting limits. This contrasts with the recent band 
structure calculations \cite{ot}, which revealed the fact that
the on-site Coulomb repulsion magnitude in these systems is 
actually relatively high, and may even reach 10eV causing 
strong correlation effects.  Thus, exploring possibilities
for other techniques compatible with strong correlation effects
is needed for describing the origin of ferromagnetism in these
systems.

In this Letter we present the result of an {\sl{exact}} 
calculation applied to a multiband Hubbard model, by which it 
can be shown rigorously that ferromagnetic ground state 
appears well below quarter filling. A similar method was 
used previously \cite{hat} in polythiophene type structures
in the high doping limit, i.e., well over the half filled case. 
In the present case however, working well below half filling,
our results are much broader than the flat band 
approach \cite{harom,haroma} because we obtain ferromagnetism 
with sparse and even random presence of the 
local Coulomb repulsion, i.e., a Hubbard term is not required on
every site. 

Our analytical approach proceeds in three steps: the transformation of 
the Hamiltonian
into positive semidefinite form, the construction of the ground state and the 
proof of their uniqueness. The technique is independent of the spatial 
dimension and does not require integrability of the model. The method
was previously applied to construct the exact ground states for the Hubbard 
chains with different geometrical structures \cite{17} and even for 
two- \cite{2DPAM} and three-dimensional periodic Anderson model 
\cite{3DPAM}. Details of the method are described in Ref. \cite{review}.

Following through with these steps, we start with {\sl{i) transformation
of a Hamiltonian into positive semidefinite form}}. A positive semidefinite 
operator $\hat P$ is defined as an
operator which has only non-negative expectation values
with all components $|\chi\rangle \in { \cal{H}}$ of the
Hilbert space ${\cal{H}}$, i.e. $\langle \chi |\hat P |\chi
\rangle \geq 0$. However, any Hamiltonian, $\hat H$, which 
describes a physical system is always bounded below and hence, can be written 
as $\hat H = \hat P + C$, where $\hat P$ a positive semidefinite and
$C$ a scalar. It follows that the most general wave vector $|\Psi_g \rangle$ 
satisfying $\hat P |\Psi_g \rangle =0$ is the ground state wave vector of 
$\hat H$ with ground state energy $E_g = C$. 

To show how straightforward the method is, we exemplify it's application
with the two-dimenisonal Hubbard model, given by the well-known Hamiltonian:
\begin{equation}
\hat H= \sum_{{\bf i},\sigma} (t_x \hat c^{\dagger}_{
{\bf i}+{\bf x},\sigma} \hat c_{{\bf i},\sigma} + 
t_y \hat c^{\dagger}_{{\bf i}+{\bf y},\sigma}
\hat c_{{\bf i},\sigma} + H.c ) + \hat H_U, \quad
\hat H_U= U \sum_{\bf i} 
\hat n_{{\bf i},\sigma} \hat n_{{\bf i},-\sigma} ,
\label{E3}
\end{equation}
on a square lattice with Bravais vectors ${\bf x},{\bf y}$ and periodic
boundary conditions. Here we used standard notations, where
$\hat c^{\dagger}_{{\bf j},\sigma}$ is the electron creation operator
with spin projection $\sigma$, $\hat n_{{\bf j},\sigma}=\hat c^{\dagger}_{
{\bf j},\sigma} \hat c_{{\bf j},\sigma}$ is the number operator, 
$t_x$ and $t_y$ are the hopping matrix elements connecting nearest 
neighbour lattice sites in $x$ and $y$ directions, and $U > 0$ 
represents the strength of the on-site Coulomb repulsion. 

Per definition $\hat H_U$ is a positive semidefinite operator, however
the total Hamiltonian, $\hat H$ is not. Hence, we perform a transformation
on $\hat H$ to obtain one. For this, 
to each square, with coordinates 
${\bf i},{\bf i}+{\bf x}, {\bf i}+{\bf x}+{\bf y},{\bf i}+{\bf y}$, 
we attach two block operators
\begin{equation}
\hat A_{{\bf i},\sigma} = a_2 \hat c_{{\bf i}+{\bf x},
\sigma} + 
a_3 \hat c_{{\bf i}+{\bf x}+{\bf y},\sigma} +
a_4 \hat c_{{\bf i}+{\bf y},\sigma}, \: \: \: \:
\hat B_{{\bf i},\sigma} = b_1 \hat c_{{\bf i},\sigma} + 
b_2 \hat c_{{\bf i}+{\bf x},\sigma} + b_4 \hat c_{{\bf i}
+{\bf y},\sigma},
\label{E4}
\end{equation}
so as the starting Hamiltonian, Eq.(\ref{E3}), transforms into
\begin{equation}
\hat H_{AB}=
\sum_{{\bf i},\sigma} ( \hat A^{\dagger}_{{\bf i},\sigma} 
\hat A_{{\bf i},\sigma}
+ \hat B^{\dagger}_{{\bf i},\sigma} \hat B_{{\bf i},
\sigma}) =\hat H - \hat H_U + q \hat N, 
\label{E5}
\end{equation}
$\hat P=\hat H_{AB}+\hat H_U$, $\quad C=-q N$,
with the number of electrons, $N$, fixed. The obtained
$\hat H_{AB}$ is a positive semidefinite operator, and 
hence $\hat P = \hat H_{AB} + \hat H_U$ also. The only task
left is to calculate the coefficients of the block operators,
$\hat A_{{\bf i},\sigma}$ and $\hat B_{{\bf i},\sigma}$, for
which the transformation into Eq.(\ref{E5}) gives:
\begin{eqnarray}
&&t_x=b_2^* b_1 + a_3^* a_4, \: \: t_y=b_4^* b_1 + a_3^* a_2, \: \:
t_{y+x}=t_{y-x}=b_4^* b_2 + a_4^* a_2=0,
\nonumber \\
&&q=|b_1|^2+|b_2|^2+|b_4|^2 +|a_2|^2 +
|a_3|^2 +|a_4|^2 .
\label{E6}
\end{eqnarray}
This system of equations represents the {\it matching conditions}. 
Obtaining a solution for these 
matching conditions implies a solution for the Hubbard Hamiltonian. 
This is usually not an easy task, as these equations are 
coupled, complex algebraic non-linear equations, but it can be
done in some restricted $\hat H$ parameter space, e.g., see,  
Refs. \cite{17,2DPAM,3DPAM} and even in disordered systems \cite{disorder}. 

Having a solution for the matching equations, we can easily 
go to the second step in our approach, namely {\sl{ii) 
the construction of the ground state}}, i.e., $|\Psi_g\rangle$. 
The solution will depend on the structure of $\hat P$, however 
the most common case is when $\hat P$ operator contains terms of the form 
$\sum_{{\bf i},\sigma} \hat A^{\dagger}_{{\bf i},\sigma} 
\hat A_{{\bf i},\sigma}$, $\sum_{{\bf i},\sigma} \hat B^{
\dagger}_{{\bf i},\sigma} \hat B_{{\bf i},\sigma}$. In these cases
the ground state is constructed with the help of a block operator 
$\hat C^{\dagger}_{{\bf j},\sigma}$ which anticommutes with
$\hat A^{}_{{\bf i},\sigma}$ and $\hat B^{}_{{\bf i},\sigma}$, 
i.e., $\{\hat A_{{\bf i},\sigma}, \hat C^{\dagger}_{{\bf j},
\sigma'} \}$ = $\{\hat B_{{\bf i},\sigma}, \hat C^{\dagger}_{
{\bf j},\sigma'} \}$ = $0$, 
for all possible values of all indices. Namely, if  
$|\chi\rangle = \prod_{{\bf i}, \sigma}\hat C^{\dagger}_{{\bf i},\sigma} |0\rangle$ 
where $|0\rangle$ is the bare vacuum, then 
$\hat H_{AB}|\chi\rangle = 0$. In order however, to obtain the ground state
$|\Psi_g\rangle$ of $\hat H$ we need $| \chi \rangle$ to control 
all positive semidefinite operators in $\hat P$. In other words 
$| \chi \rangle$ has to be inserted in the kernel \cite{kernel} of 
all positive semidefinite operators existing in $\hat P$. 
This process is easily implemented by
imposing some restrictions 
$({\bf i},\sigma) \in {\cal{M}}$ on the validity domain 
of $|\chi\rangle$, to be determined separately on model basis, 
after which the true ground state becomes 
$|\Psi_g\rangle = \prod_{({\bf i},
\sigma)\in {\cal{M}}}\hat C^{\dagger}_{{\bf i},\sigma} 
|0\rangle$. 

The last step in our approach is {\sl{iii) the 
proof of their uniqueness}}. 
For the most general case, when the 
ground state $|\Psi_g(m)\rangle$ is $M$ fold
degenerate (i.e. $m=1,2,...,M$), the proof of the
uniqueness is done in two steps. In the first step, 
we prove 
that for all possible $m$ values $|\Psi_g(m)\rangle \in
Ker(\hat P)$ is true. In the second step, we verify that any
arbitrary wave vector $|\nu\rangle \in Ker(\hat P)$ can be
expressed as a linear combination of the $|\Psi_g(m)\rangle$
terms, see Ref. \cite{3DPAM,review,ujreview}. In the 
non-degenerate case the steps are the same, but applied only
to the $m=1$ ground state component. 

\begin{figure}[t]
\vspace{40pt}
\begin{center}
\includegraphics{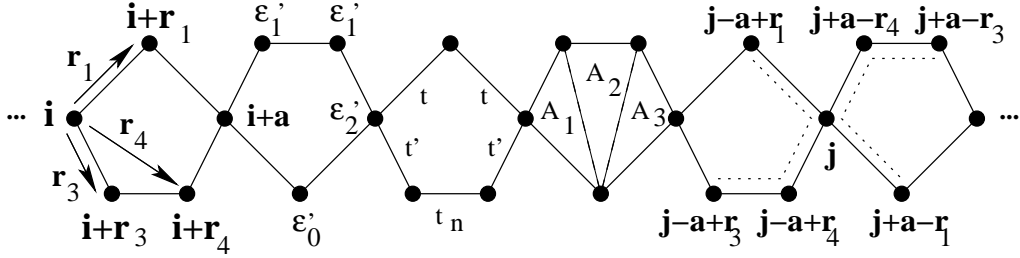}
\caption{The pentagon chain without external links. The first pentagon 
at site {\bf{i}} shows the site coordinates; 
the on-site potentials and the hopping transfer mattrixes are
shown on the second and third pentagons, respectively. 
The fourth pentagon depicts the triangular regions  
on which the block operators of Eq. (7) are defined. Finally, the
last two pentagons present the sites (connected with dotted line 
as a guide to the eye) which contribute to the block operator 
$\hat B^{\dagger}_{\bf{j}, \sigma}$ defined in Eq. (12). 
}
\label{figone}
\end{center}
\end{figure}

In the following, we apply the above method to two cases of pentagon chains.
First we analyse the pentagon chain without external links, see Fig. 
(\ref{figone}). 
This system is a conductor, a conjugated polymer of great interest 
which has not been analyzed yet with rigurous methods only 
way above the half filled concentration regime. 
Each pentagon cell contains four sites per cell. 
The cell defined at any site ${\bf i}$, see, the first cell of Fig. 
(\ref{figone}), 
has four adjacent sites at ${\bf i}+{\bf r}_n$, where $n=1,2,3,4$, and 
by convention
${\bf r}_2=0$. For a fixed $n$, the sites ${\bf i}+{\bf r}_n$ are belonging to 
the $n$-th sublattice. The on-site potentials and hopping transfer 
we used the notations shown on the second and third cell of the pentagon 
chain of Fig. (\ref{figone}). 

With the above notations, 
the non-interacting part of the Hamiltonian becomes
$
\hat H_0 = \sum_{\sigma} \sum_{{\bf i}=1}^{N_c} \: \{ \: [ \: t ( 
\hat c^{\dagger}_{{\bf i}+{\bf r}_1,\sigma} \hat c_{{\bf i},\sigma} +
\hat c^{\dagger}_{{\bf i}+{\bf a},\sigma} \hat c_{{\bf i}+{\bf r}_1,\sigma}) +
t_n \hat c^{\dagger}_{{\bf i}+{\bf r}_3,\sigma} \hat c_{{\bf i}+{\bf r}_4,
\sigma} 
+ t' ( 
\hat c^{\dagger}_{{\bf i},\sigma} \hat c_{{\bf i} +{\bf r}_3,\sigma} +
\hat c^{\dagger}_{{\bf i}+{\bf r}_4,\sigma} \hat c_{{\bf i}+{\bf a},\sigma}) 
+ H.c. ]
+ \epsilon'_0 \hat n_{{\bf i}+{\bf r}_1,\sigma} +
+ \epsilon'_1 (\hat n_{{\bf i}+{\bf r}_3,\sigma} + 
\hat n_{{\bf i}+{\bf r}_4,\sigma}) + \epsilon'_2 \hat n_{{\bf i},\sigma}) 
\} ,
$
where $N_c$ represents the number of cells. There are 
$4 N_c$ lattice sites in the system and $N$ electrons.

While, the interacting part of the Hamiltonian is 
$
\hat H_U = \sum_{{\bf i}=1}^{N_c} [
U_0 \hat n_{{\bf i}+{\bf r}_1,\uparrow} \hat n_{{\bf i}+{\bf r}_1,\downarrow}+
U_1 (\hat n_{{\bf i}+{\bf r}_3,\uparrow} \hat n_{{\bf i}+{\bf r}_3,\downarrow}+
\hat n_{{\bf i}+{\bf r}_4,\uparrow} \hat n_{{\bf i}+{\bf r}_4,\downarrow}) +
U_2 \hat n_{{\bf i},\uparrow} \hat n_{{\bf i},\downarrow}],
$
where, since in the positions ${\bf i}+{\bf r}_1$,
$({\bf i}+{\bf r}_3,{\bf i}+{\bf r}_4)$, and ${\bf i}$
different type of atoms are potentially present holding in order the 
on-site one-particle potentials
$\epsilon'_0, \epsilon'_1$ and $\epsilon'_2$,  three different 
$U_0,U_1,U_2 >0$ on-site Coulomb repulsion (Hubbard interaction) values are 
used. One has the Hubbard
$U_n$ at the site where the on-site potential is $\epsilon'_n$.

Hence, the total Hamiltonian will be simply $\hat H = \hat H_0 + \hat H_U$
and using the technique previously detailed, see, 
Eqs. (\ref{E4}) - ({\ref{E6}), the 
block operators are defined as: 
\begin{eqnarray}
&&\hat A_{1,{\bf i},\sigma}= a_{1,1} \hat c_{{\bf i}+{\bf r}_1,\sigma} +
a_{1,2} \hat c_{{\bf i},\sigma} + a_{1,3} \hat c_{{\bf i}+{\bf r}_3,\sigma} ,
\nonumber\\
&&\hat A_{2,{\bf i},\sigma}= a_{2,1} \hat c_{{\bf i}+{\bf r}_1,\sigma} +
a_{2,3} \hat c_{{\bf i}+{\bf r}_3,\sigma} + a_{2,4} \hat c_{{\bf i}+{\bf r}_4,
\sigma} ,
\nonumber\\  
&&\hat A_{3,{\bf i},\sigma}= a_{3,1} \hat c_{{\bf i}+{\bf r}_1,\sigma} +
a_{3,4} \hat c_{{\bf i}+{\bf r}_4,\sigma} + a_{3,5} \hat c_{{\bf i}+{\bf a},
\sigma} .
\label{s2.14}
\end{eqnarray}
These operators span \cite{otvertex} a pentagon cell as depicted in 
the fourth cell
of Fig.(\ref{figone}). Using periodic boundary conditions $\hat H_0$ 
transfroms into:
\begin{eqnarray}
\hat H_0 = \sum_{\sigma} \sum_{{\bf i}=1}^{N_c} \sum_{m=1}^3
\hat A^{\dagger}_{m,{\bf i},\sigma} \hat A_{m,{\bf i},\sigma} .
\label{s2.15}
\end{eqnarray}
We are interested to find the ground state
solution well below quarter filling, hence we work in the condition 
$N \le N_c$. 
For the solution of the matching conditions, 
for real hopping matrix elements and conditions
$t_n > 0, \epsilon'_1 - t_n > 0 $ we obtained: 
\begin{eqnarray}
&&a_{1,1}=e^{i \phi_1} |a_{1,1}|, \quad  a_{1,2}= e^{i \phi_1} \frac{t}{
|a_{1,1}|}, \quad a_{1,3}= e^{i \phi_1} \frac{t'}{t} |a_{1,1}|, 
\nonumber\\
&&a_{2,1}=e^{i \phi_2} |a_{2,1}|, \quad a_{2,3}=-e^{i \phi_2} \frac{t'}{t}
\frac{|a_{1,1}|^2}{|a_{2,1}|}, \quad a_{2,4}=-e^{i \phi_2} \frac{t t_n}{t'}
\frac{|a_{2,1}|}{|a_{1,1}|^2},     
\nonumber\\
&&a_{3,1}= e^{i \phi_3}|a_{3,1}|, \quad a_{3,4}= e^{i \phi_3} \frac{t'}{t}
|a_{3,1}|, \quad a_{3,5}= e^{i \phi_3} \frac{t}{|a_{3,1}|} ,
\label{s2.18}
\end{eqnarray}
where $\phi_{m}$, $m=1,2,3$ are arbitrary phases. In Eqs. (\ref{s2.18}) 
the Hamiltonian parameters $t,t',t_n,\epsilon'_1$ can be arbitrary chosen,
while $\epsilon'_0, \epsilon'_2$ are given by the conditions
$\epsilon'_0= [t^2/(t'^2 t_n)](\epsilon'^2_1 -t^2_n)$, and
$\epsilon'_2=2 t'^2/(\epsilon'_1-t_n)$. These last two conditions provide
the lowest flat band of the band structure. 

Since $\hat H_0$ has the simple expression (\ref{s2.15}), we look for 
the ground state wave function in the form
\begin{eqnarray}
|\Psi_g\rangle = \prod_{{\bf i}=1}^{N \leq N_c} \hat B^{\dagger}_{{\bf i},
\sigma_{\bf i}} |0\rangle ,
\label{s2.20}
\end{eqnarray}
where $|0\rangle$ is the bare vacuum, and $\hat B^{\dagger}_{{\bf i},
\sigma_{\bf i}}$ satisfies for all $n=1,2,3$ the relation
\begin{eqnarray}
\{ \hat A_{n,{\bf i},\sigma}, \hat B^{\dagger}_{{\bf i}',\sigma'_{{\bf i}'}}
\} =0,
\label{s2.21}
\end{eqnarray}
where ${\bf i},{\bf i}',\sigma, \sigma'_{{\bf i}'}$ are arbitrary. 
Since only one type of canonical Fermi operator is defined on each site, 
Eq. (\ref{s2.20}) is true 
if the $\hat B^{\dagger}_{{\bf i}',\sigma'_{{\bf i}'}}$ operators 
do not overlap, or the neighbouring operators overlap at least on one site. 

The first case, when the $\hat B^{\dagger}_{{\bf i}',\sigma'_{{\bf i}'}}$ 
operators do not overlap, would mean 
a localized and paramagnetic ground state of the general form 
$\hat B^{\dagger}_{{\bf i},\sigma} = x_1 \hat c^{\dagger}_{{\bf i}+{\bf r}_1,
\sigma} + x_3 \hat c^{\dagger}_{{\bf i}+{\bf r}_3,\sigma} + x_4 
\hat c^{\dagger}_{{\bf i}+{\bf r}_4,\sigma}$. However, there isn't any
value of $x_1, x_2, x_3$, except $x_1=x_3=x_4=0$, which would satisfy
Eq. (\ref{s2.21}), hence there is no solution in this case.

To search for a solution in the second case, i.e., when the
$\hat B^{\dagger}_{{\bf i}',\sigma'_{{\bf i}'}}$ operators overlap, we
define $\hat B^{\dagger}_{{\bf i},\sigma}$ as shown on the last two cells
of Fig. (\ref{figone}), namely:
\begin{eqnarray}
\hat B^{\dagger}_{{\bf i},\sigma} &=& x_1 \hat c^{\dagger}_{{\bf i}+{\bf r}_1,
\sigma} + x_2 \hat c^{\dagger}_{{\bf i},\sigma} + x_3 
\hat c^{\dagger}_{{\bf i}+{\bf r}_3,\sigma} + x_4 \hat c^{\dagger}_{{\bf i}
+{\bf r}_4,\sigma}  
\nonumber\\
&+& y_1 \hat c^{\dagger}_{{\bf i}-{\bf a}+{\bf r}_1,\sigma} + 
y_3 \hat c^{\dagger}_{{\bf i}-{\bf a}+{\bf r}_3,\sigma} + y_4 
\hat c^{\dagger}_{{\bf i}-{\bf a}+{\bf r}_4,\sigma} ,
\label{s2.23}
\end{eqnarray}
and the solution to (\ref{s2.21}) is: 
\begin{eqnarray}
&&x_4=-\frac{t}{t'} x_1, \quad x_3 =\frac{t\epsilon'_1}{t' t_n} x_1, 
\quad x_2=- \frac{t}{t'^2 t_n} 
(\epsilon'^2_1-t^2_n)  x_1, 
\nonumber\\
&&y_1=x_1, \quad y_3=x_4=-\frac{t}{t'} x_1, \quad y_4=x_3= \frac{
t\epsilon'_1}{t' t_n} x_1 . 
\label{s2.26}
\end{eqnarray}
Consequently, the $\hat B^{\dagger}_{{\bf i},\sigma}$ operator becomes
\begin{eqnarray}
\hat B^{\dagger}_{{\bf i},\sigma} &=& x_1 [
- \frac{t(\epsilon'^2_1-t^2_n)}{t'^2 t_n} \hat c^{\dagger}_{{\bf i},\sigma} 
+ (\hat c^{\dagger}_{{\bf i}+{\bf r}_1,\sigma} + \hat c^{\dagger}_{{\bf i}
-{\bf a}+{\bf r}_1,\sigma}) + \frac{t\epsilon'_1}{t't_n} ( \hat c^{\dagger}_{
{\bf i}+{\bf r}_3,\sigma} + \hat c^{\dagger}_{{\bf i}-{\bf a}+{\bf r}_3,
\sigma}) 
\nonumber\\
&-& \frac{t}{t'} ( \hat c^{\dagger}_{{\bf i}+{\bf r}_4,\sigma} 
+ \hat c^{\dagger}_{{\bf i}-{\bf a}+{\bf r}_3,\sigma} ) ] . 
\label{s2.27}
\end{eqnarray}

The (unnormalized) ground state wave function at $1/8$ filling (e.g. $N = N_c$)
becomes a saturated ferromagnet
\begin{eqnarray}
|\Psi_g\rangle = \prod_{{\bf i}=1}^{N_c} \hat B^{\dagger}_{{\bf i},\sigma}
|0\rangle ,
\label{s2.28}
\end{eqnarray}
where $\sigma$ is fixed. Below $1/8$ filling the block
operator $\hat B^{\dagger}_{{\bf i}, \sigma}$ is still given in Eq. (\ref{s2.27}), 
and the ground state will have the (\ref{s2.20}) form. But, a
geometrical degeneracy occurs: only overlaping $\hat B^{\dagger}_{{\bf i},
\sigma}$ operators will have the same spin index. Hence, the ground state 
will be composed from ferromagnetic clusters which if don't overlap,
will have arbitrary spin orientations. We should also point out that
Eq. (\ref{s2.28}) corresponds to the half filled lower flat band. The 
obtained solution is true for arbitrary large
$U_0,U_1,U_2 > 0$ Hubbard terms. 

Next, we analyse the second model of a pentagon chain,
namely the pentagon chain with external links and antennas, see 
Fig. (\ref{figtwo}).   
This chain is also a conductor, and we are going to show in the following 
that the obtained results are qualitatively the same as in the previous
case. The new pentagon chain, with external links and antennas connected to
the pentagons, is shown in Fig. (\ref{figtwo}). The cell
now contains six sites and consequently, there will be six 
sublattices in the system. The cell defined at any site 
${\bf i}$, see, the first cell of Fig. (2) has six adjacent sites at
${\bf i}+{\bf r}_n$, where now $n=1,2,...,6$, and ${\bf r}_3=0$ 
by convention. For a fixed $n$, the sites ${\bf i}+{\bf r}_n$ are 
belonging to the $n$-th sublattice. 

\begin{figure}[t]
\vspace{40pt}
\begin{center}
\includegraphics{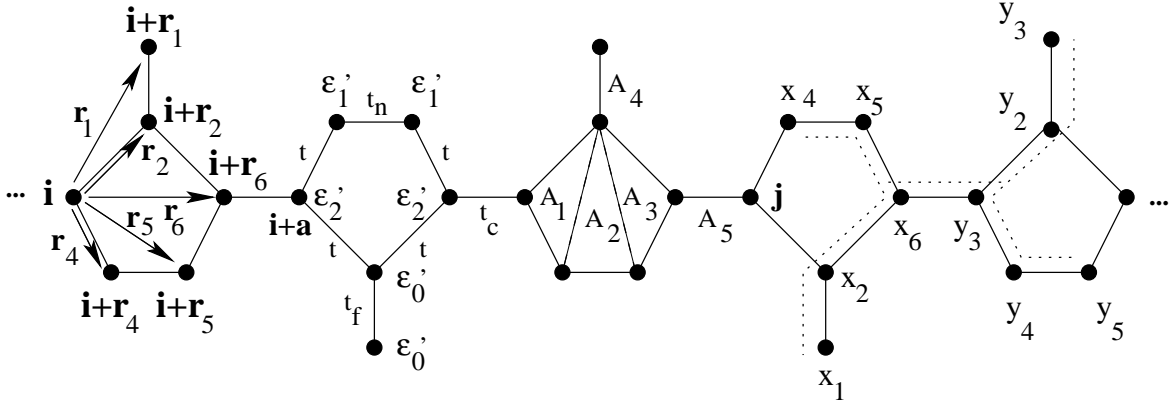}
\caption{The pentagon chain with external links and antennas. 
The first pentagon 
at site {\bf{i}} shows the site coordinates; the on-site potentials and the 
hopping transfermattrixes are shown on the second pentagon. 
The third pentagon depicts the five regions 
(three triangular and two bond domains) on which the block 
operators from Eq. (18) are defined. Finally, the last two pentagons present
the sites (connected with dotted line as a guide to the eye) which 
contribute to the block operator 
$\hat B^{dagger}_{{\bf j}, \sigma}$ defined in 
Eq. (23). The coefficients of each individual site
contributions ($x_i$, $y_i$) are also shown. 
}
\label{figtwo}
\end{center}
\end{figure}

With the on-site potentials and hopping matrix elements defined on the second
and third cell of Fig. (2), the starting Hamiltonian 
$\hat H= \hat H_0 + \hat H_U$ becomes:
$
\hat H_0 = \sum_{\sigma} \sum_{{\bf i}=1}^{N_c} \: \{ \: [ \: t_f 
\hat c^{\dagger}_{{\bf i}+{\bf r}_1,\sigma} \hat c_{{\bf i}+{\bf r}_2,\sigma} +
t_c \hat c^{\dagger}_{{\bf i}+{\bf r}_6,\sigma} \hat c_{{\bf i}+{\bf a},
\sigma} + t_n \hat c^{\dagger}_{{\bf i}+{\bf r}_4,\sigma} \hat c_{{\bf i}+
{\bf r}_5,\sigma} 
+ t ( 
\hat c^{\dagger}_{{\bf i}+{\bf r}_2,\sigma} \hat c_{{\bf i},\sigma} +
\hat c^{\dagger}_{{\bf i},\sigma} \hat c_{{\bf i}+{\bf r}_4,\sigma} +
\hat c^{\dagger}_{{\bf i}+{\bf r}_5,\sigma}\hat c_{{\bf i}+{\bf r}_6,\sigma} +
\hat c^{\dagger}_{{\bf i}+{\bf r}_6,\sigma}\hat c_{{\bf i}+{\bf r}_2,\sigma} 
) + H.c. ]
+ \epsilon'_0 (\hat n_{{\bf i}+{\bf r}_1,\sigma} +
\hat n_{{\bf i}+{\bf r}_2,\sigma}) + \epsilon'_1
(\hat n_{{\bf i}+{\bf r}_4,\sigma} + \hat n_{{\bf i}+{\bf r}_5,\sigma}) +
\epsilon'_2 (\hat n_{{\bf i},\sigma} + \hat n_{{\bf i}+{\bf r}_6,\sigma})
\},
$
while the interacting part of the Hamiltonian is now 
$
\hat H_U = \sum_{{\bf i}=1}^{N_c} [
U_0 (\hat n_{{\bf i}+{\bf r}_1,\uparrow} \hat n_{{\bf i}+{\bf r}_1,\downarrow}+
\hat n_{{\bf i}+{\bf r}_2,\uparrow} \hat n_{{\bf i}+{\bf r}_2,\downarrow})+
U_1 (\hat n_{{\bf i}+{\bf r}_4,\uparrow} \hat n_{{\bf i}+{\bf r}_4,\downarrow}+
\hat n_{{\bf i}+{\bf r}_5,\uparrow} \hat n_{{\bf i}+{\bf r}_5,\downarrow}) 
+
U_2 (\hat n_{{\bf i},\uparrow} \hat n_{{\bf i},\downarrow}+
\hat n_{{\bf i}+{\bf r}_6,\uparrow} \hat n_{{\bf i}+{\bf r}_6,\downarrow}) ].
$
In the interacting part of the Hamiltonian,
since in positions $({\bf i}+{\bf r}_1,{\bf i}+{\bf r}_2)$,
$({\bf i}+{\bf r}_4,{\bf i}+{\bf r}_5)$, and $({\bf i},{\bf i}+{\bf r}_6)$
different type of atoms are present, three are three different 
$U_0,U_1,U_2 >0$ local Coulomb repulsion values. 
Also, the number of lattice sites on the chain is $6 N_c$, 
and the number of electrons is $N$.

Using the same approach as for the previously analysed case, we
obtain for the 
block operator $\hat B^{\dagger}_{{\bf i},\sigma}$:
\begin{eqnarray}
\hat B^{\dagger}_{{\bf i},\sigma} &=& x_1 [\hat c^{\dagger}_{{\bf i}+{\bf r}_1,
\sigma} - \frac{\epsilon'_0}{t_f}\hat c^{\dagger}_{{\bf i}+{\bf r}_2,\sigma} 
+ \frac{\epsilon'_0}{t_f} \hat c^{\dagger}_{{\bf i}+{\bf r}_4,\sigma} 
- \frac{\epsilon'_0 \epsilon'_1}{t_f t_n} \hat c^{\dagger}_{{\bf i}
+{\bf r}_5,\sigma} + \frac{\epsilon'_0(\epsilon'^2_1-t^2_n)}{t t_f t_n} 
 \hat c^{\dagger}_{{\bf i}+{\bf r}_6,\sigma} 
\nonumber\\
&-&(\epsilon'_1-t_n) sign(t_c) \: \: ( \: \:
\hat c^{\dagger}_{{\bf i}+{\bf a}+{\bf r}_1,\sigma} - 
\frac{\epsilon'_0}{t_f} \hat c^{\dagger}_{{\bf i}+{\bf a}+{\bf r}_2,\sigma} + 
\frac{\epsilon'_0}{t_f}  \hat c^{\dagger}_{{\bf i}+{\bf a}+{\bf r}_5,\sigma}
\nonumber\\
&-&
\frac{\epsilon'_0 \epsilon'_1}{t_ft_n} \hat c^{\dagger}_{{\bf i}+{\bf a}+
{\bf r}_4,\sigma} + 
\frac{\epsilon'_0(\epsilon'^2_1 - t^2_n)}{t t_f t_n} 
\hat c^{\dagger}_{{\bf i}+{\bf a},\sigma} \: \: ) \: \: ] .
\label{ss2.26}
\end{eqnarray}

The (unnormalized) ground state wave function at $1/12$ filling (e.g. $N = N_c$)
becomes a saturated ferromagnet
\begin{eqnarray}
|\Psi_g\rangle = \prod_{{\bf i}=1}^{N_c} \hat B^{\dagger}_{{\bf i},\sigma}
|0\rangle ,
\label{ss2.27}
\end{eqnarray}
where $\sigma$ is fixed. Below $1/12$ filling the expression of 
$\hat B^{\dagger}_{{\bf i},\sigma}$ remains as given in Eq. (\ref{ss2.26}), 
but in (\ref{ss2.27}) a geometrical degeneracy occurs, only overlaping 
$\hat B^{\dagger}_{{\bf i},\sigma}$ operators will have the same spin 
index, and the ground state will be constructed from ferromagnetic clusters 
which if not in contact, will have arbitrary spin orientation. 
The ground state given by Eq. (\ref{ss2.27}) corresponds to a 
half filled lowest flat band. The obtained solution however, occurs 
for arbitrary large $U_0,U_1,U_2 > 0$ Hubbard interaction terms. 
Similar situations for other compounds have been intensively analyzed in
literature \cite{Z1,Z2,Z3}.

In summary, by employing a rigorous analytical method we have constructed
exact ground states for multiorbital pentagon Hubbard chains. The 
ferromagnetism what we found well below half filling originate from 
the multi-orbital polygon chains which yield dispersionless band
in the presence of site-dependent Coulomb
intercation. $N$ dependent ground states we have obtained 
for $N \leq N_c$, and the system is conducting for $N < N_c$.
At $N=N_c$, the ferromagnetism emerges since in the ground state 
wave vector all contributing terms have the same fixed spin projection. 
The proof of the uniqueness of our results 
can be made along the lines of Refs. \cite{review,ujreview}. 

Organic ferromagnets have attracted much attention  
as a challenging target. In particular, organic magnets consisting entirely
of non-magnetic elements is of fundamental as well as practical interest. 
Ordinary ferromagnets consist of magnetic elements and even 
one-dimensionals models which exhibit ferromagnetism exploit
electrons in $d$ or $f$ orbitals. In the presence of strong interaction, 
for example such in the Kondo lattice case, the $f$ electrons are responsible
for ferromagnetism which, as it was shown in Refs. \cite{klm} using
non-Abelian density matrix renormalization group \cite{dmrg}, 
order due to scattering with the conduction 
electrons. Since hopping is energetically most favorable for
conductions electrons which preserver their spins, called coherent
hopping, this tends to align the localized $f$ electron spins \cite{klm}. 

But, in the cases analysed in this Letter only non-magnetic elements
are present in the pentagon chain. We can rightfully ask the question
how magnetism can arise in these systems? 
The answer to this question is that the Coulomb intercations are 
capable of turning itinerant system into a ferromagnetic phase
in an extended parameter region. 
The magnetism arises as an effect of the electron-electron repulsion 
when the adjacent block operators which yield the ground state wave vector
overlap and intuitively the spin has to
align to lower the repulsion energy due to Pauli's principle. 

Continuing the above agurment, due to the overlaping adjacent block operators, 
in our model we do not even need all sites to be interacting, it is enoguh
to have merely one site to be intercating in each cell. 
To show this, let us consider first 
the pentagon chain wihtout external links. The sites contributing
to the block operators of the ground state wave vector are shown in
Fig. (\ref{figone}) with dotted lines (last two pentagon of the figure).
Consequently, at $N=N_c$ number of electrons, ferromagnetism will appear 
even if there is only one site in each pentagon with non zero
local Coulomb repulsion, namely one of the sites with coordinates
${\bf r}_1, {\bf r}_3, {\bf r}_4$. On these sites, even
a random distribution of one local Coulomb repulsion on each cell yields 
ferromagnetism. We note that in this case 75\% of sites are 
non-interacting (three sites from four in each cell), i.e. without Hubbard 
interaction. 

The same is true for the case of pentagon chains with 
external links. If one site per pentagon has a Hubbard U attached to it, 
in between sites with coordinates ${\bf r}_1, {\bf r}_2, {\bf r}_4,
{\bf r}_5, {\bf r}_6$, see, Fig. (\ref{figtwo}). 
In this case 83.3\% of sites are 
non-interacting (five sites out of six in each cell). 
This shows that surprizingly, the complete absence of magnetic atoms 
with sparse and even random presence of the 
local Coulomb repulsion can lead to ferromagnetism. This underlines that the
conditions in which we obtained ferromagnetism are 
much broader than those fixed by flat-band ferromagnetism, 
where on every site of the system $U>0$ is required \cite{Z4}. Hence, 
the obtained solutions point to a new route for the design
of ferromagnetic chain polymners.

Regarding the experimental observation of ferromagnetism, we have
to point out that the required electron doping of the pentagon chains
can be achieved \cite{21} by changing the Fermi level by selecting
appropriate side groups or by field-effect doping in a double-layer
transisitor structure \cite{22}. Indeed, depending on the applied doping levels
pentagon polymers can be turned \cite{2324} into ferromagnets, 
spin glasses or simple paramagnetic polymers. 

\subsection*{Acknowledgements}
\begin{enumerate}
\item[(1)]For M. Gul\'acsi  
this research was realized in the frames of TAMOP 4.2.4. A/2-11-1-2012-0001 
"National Excellence Program - Elaborating and operating an inland student 
and researcher personal support system". The project was subsidized by the 
European Union and co-financed by the European Social Fund. 
\item[(2)]Zs. Gul\'acsi kindly acknowledges financial support provided by Alexander von 
Humboldt Foundation, OTKA-K-100288 (Hungarian Research Funds for Basic 
Research) and TAMOP 4.2.2/A-11/1/KONV-2012-0036 (co-financed by EU and European 
Social Fund). 
\end{enumerate}



\begin{thebibliography}{10}
\bibitem[14]{kernel}
The kernel $Ker (\hat O)$ of an arbitrary operator
$\hat O$ is a Hilbert subspace containing all wave vectors 
$| \phi \rangle$ with the property $\hat O | \phi \rangle = 0$.
\end{thebibliography}
\end{document}